\input harvmac
\input epsf

\def\frac#1#2{{#1\over #2}}

\newcount\figno
\figno=0 
\def\fig#1#2#3{
\par\begingroup\parindent=0pt\leftskip=1cm\rightskip=1cm\parindent=0pt
\baselineskip=11pt
\global\advance\figno by 1
\midinsert
\epsfxsize=#3
\centerline{\epsfbox{#2}}
\vskip 12pt
{\bf Fig.\ \the\figno: } #1\par
\endinsert\endgroup\par
}
\def\figlabel#1{\xdef#1{\the\figno}}
\noblackbox
\lref\BMN{
D.~Berenstein, J.~M.~Maldacena and H.~Nastase,
``Strings in flat space and pp waves from N = 4 super Yang Mills,''
JHEP {\bf 0204}, 013 (2002)
[arXiv:hep-th/0202021].
}

\lref\GrossSU{
D.~J.~Gross, A.~Mikhailov and R.~Roiban,
``Operators with large R charge in N = 4 Yang-Mills theory,''
Annals Phys.\  {\bf 301}, 31 (2002)
[arXiv:hep-th/0205066].
}

\lref\SantambrogioSB{
A.~Santambrogio and D.~Zanon,
``Exact anomalous dimensions of N = 4 Yang-Mills operators with large R
charge,''
Phys.\ Lett.\ B {\bf 545}, 425 (2002)
[arXiv:hep-th/0206079].
}

\lref\MetsaevBJ{
R.~R.~Metsaev,
``Type IIB Green-Schwarz superstring in plane wave Ramond-Ramond  background,''
Nucl.\ Phys.\ B {\bf 625}, 70 (2002)
[arXiv:hep-th/0112044].
}

\lref\MetsaevRE{
R.~R.~Metsaev and A.~A.~Tseytlin,
``Exactly solvable model of superstring in plane wave Ramond-Ramond
background,''
Phys.\ Rev.\ D {\bf 65}, 126004 (2002)
[arXiv:hep-th/0202109].
}

\lref\gsb{
M.~B.~Green, J.~H.~Schwarz and L.~Brink,
``Superfield Theory Of Type II Superstrings,''
Nucl.\ Phys.\ B {\bf 219}, 437 (1983).
}

\lref\gsthree{
M.~B.~Green and J.~H.~Schwarz,
``Superstring Field Theory,''
Nucl.\ Phys.\ B {\bf 243}, 475 (1984).
}

\lref\svone{
M.~Spradlin and A.~Volovich,
``Superstring interactions in a pp-wave background,''
Phys.\ Rev.\ D {\bf 66}, 086004 (2002)
[arXiv:hep-th/0204146].
}

\lref\svtwo{
M.~Spradlin and A.~Volovich,
``Superstring interactions in a pp-wave background II,''
arXiv:hep-th/0206073v3.
}

\lref\semenoffone{
C.~Kristjansen, J.~Plefka, G.~W.~Semenoff and M.~Staudacher,
``A new double-scaling limit of N = 4 super Yang-Mills theory and PP-wave
strings,''
Nucl.\ Phys.\ B {\bf 643}, 3 (2002)
[arXiv:hep-th/0205033].
}

\lref\bn{
D.~Berenstein and H.~Nastase,
``On lightcone string field theory from super Yang-Mills and holography,''
arXiv:hep-th/0205048.
}

\lref\seven{
N.~R.~Constable, D.~Z.~Freedman, M.~Headrick, S.~Minwalla, L.~Motl,
A.~Postnikov and W.~Skiba,
``PP-wave string interactions from perturbative Yang-Mills theory,''
JHEP {\bf 0207}, 017 (2002)
[arXiv:hep-th/0205089].
}

\lref\herman{
H.~Verlinde,
``Bits, matrices and 1/N,''
arXiv:hep-th/0206059.
}

\lref\vv{
D.~Vaman and H.~Verlinde,
``Bit strings from N = 4 gauge theory,''
arXiv:hep-th/0209215.
}

\lref\basisone{
N.~Beisert, C.~Kristjansen, J.~Plefka, G.~W.~Semenoff and M.~Staudacher,
``BMN correlators and operator mixing in N = 4 super Yang-Mills theory,''
arXiv:hep-th/0208178.
}

\lref\chuone{
C.~S.~Chu, V.~V.~Khoze and G.~Travaglini,
``Three-point functions in N = 4 Yang-Mills theory and pp-waves,''
JHEP {\bf 0206}, 011 (2002)
[arXiv:hep-th/0206005].
}

\lref\basistwo{
D.~J.~Gross, A.~Mikhailov and R.~Roiban,
``A calculation of the plane wave string Hamiltonian from N = 4
super-Yang-Mills theory,''
arXiv:hep-th/0208231v2.
}

\lref\basisthree{
N.~R.~Constable, D.~Z.~Freedman, M.~Headrick and S.~Minwalla,
``Operator mixing and the BMN correspondence,''
JHEP {\bf 0210}, 068 (2002)
[arXiv:hep-th/0209002].
}

\lref\psvvv{
J.~Pearson, M.~Spradlin, D.~Vaman, H.~Verlinde and A.~Volovich,
``Tracing the string: BMN correspondence at finite $J^2/N$,''
arXiv:hep-th/0210102.
}

\lref\gssugra{
M.~B.~Green and J.~H.~Schwarz,
``Extended Supergravity In Ten-Dimensions,''
Phys.\ Lett.\ B {\bf 122}, 143 (1983).
}

\lref\chutwo{
C.~S.~Chu, V.~V.~Khoze, M.~Petrini, R.~Russo and A.~Tanzini,
``A note on string interaction on the pp-wave background,''
arXiv:hep-th/0208148.
}

\lref\john{
J.~H.~Schwarz,
``Comments on superstring interactions in a plane-wave background,''
JHEP {\bf 0209}, 058 (2002)
[arXiv:hep-th/0208179].
}

\lref\ari{
A.~Pankiewicz,
``More comments on superstring interactions in the pp-wave background,''
JHEP {\bf 0209}, 056 (2002)
[arXiv:hep-th/0208209].
}

\lref\aristefan{
A.~Pankiewicz and B.~Stefa\'nski, Jr.,
``PP-Wave Light-Cone Superstring Field Theory,''
arXiv:hep-th/0210246.
}

\lref\chuthree{
C.~S.~Chu, M.~Petrini, R.~Russo and A.~Tanzini,
``String interactions and discrete symmetries of the pp-wave background,''
arXiv:hep-th/0211188.
}

\lref\hssv{
Y.~H.~He, J.~H.~Schwarz, M.~Spradlin and A.~Volovich,
``Explicit Formulas for Neumann Coefficients in the Plane-Wave
Geometry,''
arXiv:hep-th/0211198.
}

\lref\ksv{
I.~R.~Klebanov, M.~Spradlin and A.~Volovich,
``New effects in gauge theory from pp-wave superstrings,''
Phys.\ Lett.\ B {\bf 548}, 111 (2002)
[arXiv:hep-th/0206221].
}

\lref\gomis{
J.~Gomis, S.~Moriyama and J.~Park,
``SYM description of SFT Hamiltonian in a pp-wave background,''
arXiv:hep-th/0210153.
}

\lref\gkone{
J.~Greensite and F.~R.~Klinkhamer,
``New Interactions For Superstrings,''
Nucl.\ Phys.\ B {\bf 281}, 269 (1987).
}

\lref\gktwo{
J.~Greensite and F.~R.~Klinkhamer,
``Contact Interactions In Closed Superstring Field Theory,''
Nucl.\ Phys.\ B {\bf 291}, 557 (1987).
}

\lref\gkthree{
J.~Greensite and F.~R.~Klinkhamer,
``Superstring Amplitudes And Contact Interactions,''
Nucl.\ Phys.\ B {\bf 304}, 108 (1988).
}

\lref\grseib{
M.~B.~Green and N.~Seiberg,
``Contact Interactions In Superstring Theory,''
Nucl.\ Phys.\ B {\bf 299}, 559 (1988).
}

\Title{\vbox{\baselineskip12pt
        \hbox{hep-th/0211220}
        \hbox{PUPT-2055}
        \hbox{NSF-ITP-02-160} 
}}{On light-cone SFT contact terms in a plane wave}

\centerline{Radu Roiban${}^{1}$,
Marcus Spradlin${}^{2}$ and Anastasia Volovich${}^{3}$
}

\bigskip

\centerline{~~~${}^{1}$~Department of Physics
~~~~~~~~~~~~~~~~~~~~~~~~~~~~~~~
${}^{2}$~Department of Physics~~~~~~}
\centerline{University of California
~~~~~~~~~~~~~~~~~~~~~~~~~~~~~~~~~~
Princeton University~~~~}
\centerline{~Santa Barbara, CA 93106
~~~~~~~~~~~~~~~~~~~~~~~~~~~~~~~~~
Princeton, NJ 08544~~~~~~
}
\centerline{\tt radu@vulcan2.physics.ucsb.edu
~~~~~~~~~~~
 spradlin@feynman.princeton.edu
}
\centerline{}
\centerline{${}^{3}$~Kavli Institute for Theoretical Physics}
\centerline{University of California}
\centerline{ Santa Barbara CA 93106}
\centerline{\tt nastja@kitp.ucsb.edu}

\vskip .3in
\centerline{\bf Abstract}

Testing the BMN correspondence at non-zero string coupling $g_s$ requires
a one-loop string field theory calculation.
At order $g_s^2$, matrix elements of the light-cone string field theory
Hamiltonian between single-string
states receive two contributions:  the iterated
cubic interaction, and a contact term $\{Q, \overline{Q}\}$
whose presence is dictated
by supersymmetry.
In this paper we calculate the leading large $\mu p^+ \alpha'$
contribution from both terms for the set of intermediate
states with two string excitations.  
We find precise agreement
with the basis-independent order $g_2^2$ results from gauge theory.

\smallskip

\Date{}

\listtoc
\writetoc

\newsec{Introduction}

Berenstein, Maldacena and Nastase have recently demonstrated \BMN\ that
there is a natural
correspondence between
the states of type IIB string theory in a plane-wave background and
a class of operators in
${\cal{N}}=4$ SU($N$) gauge theory with large R-charge $J$.
This correspondence involves a limit of the gauge theory
in which the quantities
\eqn\aaa{
\lambda' = {g_{\rm YM}^2 N \over J^2}, \qquad
g_2 = {J^2 \over N}
}
and $g_{\rm YM}^2$
are held fixed while $N$ and $J$ are taken to infinity.
Moreover, the relation
\eqn\pdj{
{2 \over \mu} P^- = \Delta - J
}
was verified
in the free theory ($g_2 = 0$), originally to order $\lambda'$
\BMN, and 
to all orders in $\lambda'$ \refs{\GrossSU,\SantambrogioSB}.
Here $P^-$ is the light-cone Hamiltonian in string theory, and
$\Delta$ is the generator of scale transformations in the gauge theory.

Since $P^-$ is a symmetry generator in a background which is not
corrected by interactions,
it is natural to expect that the relation \pdj\ continues to hold in
the presence of string interactions, as
proposed in \basistwo. 
Light-cone string field
theory in the plane wave background has been constructed
in \refs{\svone,\svtwo} and developed further in \refs{\ksv \chutwo \john 
\ari \aristefan \psvvv \chuthree - \hssv},
generalizing the flat space construction of \refs{\gsb,\gsthree}.
On the gauge theory side, a number
of papers have addressed the calculation of non-planar diagrams 
\refs{
\basistwo, \semenoffone \bn \seven \chuone \basisone \basistwo-\basisthree}.
However, it is important to note that despite some suggestive
hints, there has so far been {\it absolutely no definitive
evidence} that \pdj\ holds
for $g_2 \ne 0$. While this may seem like a strong statement, it is based
on the fact that at order $g_2$ the matching of the matrix elements on the 
two sides of equation \pdj\ fixes the state-operator map and thus cannot be 
used also as a test.

The single- and double-trace BMN operators in the gauge theory cannot
be identified with single- and double-string states at finite
$g_2$ because the former are not orthogonal \refs{\basistwo,
\basisone,\basisthree}.  Therefore we
cannot check the relation \pdj\ 
at the level of {\it matrix elements} without
first identifying the $g_2$-dependent basis transformation
between BMN operators and string states.  Instead, the BMN correspondence
only requires that ${2 \over \mu} P^-$ and $\Delta - J$ have
the same {\it eigenvalues}.
On the gauge theory side,
the eigenvalues of the anomalous dimension matrix
have been calculated to order $g_2^2$ within the subspace
of BMN operators with two scalar impurities \refs{\basisone,
\basisthree}, with the result
\eqn\evalue{
(\Delta - J)_n = 2 + \lambda' \left[
n^2 + 
{g_2^2 \over 4 \pi^2} \left({1 \over 12}
+ {35 \over 32 \pi^2 n^2}\right)\right].
}
This result was recovered in \psvvv\ using a combination
of string bit model predictions and ${\cal{O}}(g_2)$ input
from string field theory computations, but an honest
string field theory
calculation of the ${\cal{O}}(g_2^2)$ term in
\evalue\ is so far lacking.

One can also turn the problem around and construct the state-operator
map so as to ensure that the equation \pdj\ is valid at the level of 
matrix elements \basistwo. It is 
important  to stress that this approach has the same predictive power
as the one above, since the existence of an operator basis in gauge theory 
such that \pdj\ is valid at the level of matrix elements is equivalent
to the equality of eigenvalues of the two operators. The essential test of the 
proposal \pdj\ and, 
more generally, of the BMN conjecture comes from comparing the matrix 
elements of other operators once the state-operator map is
fixed.
A proposal for the state-operator map
to all orders in $g_2$
has been recently put forward in \psvvv.

In this paper we compute
matrix elements  of the string field theory
Hamiltonian $P^-$ at order $g_s^2$ 
in the limit of large $\mu p^+ \alpha'$
between two single-string states
for the same set of intermediate states used in gauge theory.
Recall that $\lambda'$ and $g_2$ are related to the
string theory parameters by $\lambda' = (\mu p^+ \alpha')^{-2}$ and
$g_2 \lambda' = 4 \pi g_s$.

The ${\cal{O}}(g_2^2)$ matrix element between single-string states is a one-loop-like 
effect, which enters
the Hamiltonian through supersymmetry via $P^-={1\over 2}\{Q, \overline{Q}
\}$.  This calculation is performed using techniques analogous to those 
used in string loop diagrams; we include only those intermediate states which 
have  one bosonic excitation and one fermionic excitation. 
These are the states corresponding to the perturbative
gauge theory calculations.

In the following section we review the procedure for determining
the state-operator correspondence at finite $g_2$ and write
the known result at order $g_2$.
In section 3 we review the large $\mu p^+ \alpha'$ limit of light-cone string
field theory in the plane wave background and calculate
the two diagrams (see Figure 1) which contribute to the eigenvalues \evalue.
We conclude with a detailed discussion of the assumptions made in our work
and some interesting questions raised by our analysis.

\newsec{BMN Correspondence at Finite $g_2$}

String field theory has a natural basis of single- (double-, etc.)
string states, and the gauge theory has a natural basis of single-
(double-, etc.) trace operators.
As mentioned in the introduction, the natural identification between
these two bases breaks down at order $g_2$ because the scalar product
is orthogonal in the string theory basis but not in the gauge theory
basis \refs{\basistwo,\basisone,\basisthree}.
The explicit transformation between 
these two bases can be worked out as follows \basistwo.
First one diagonalizes the scalar product in the gauge theory,
which fixes
the basis transformation up to an arbitrary orthogonal transformation.
This remaining ambiguity is fixed by requiring the matrix elements
of \pdj\ to agree.
At the end of the day, this procedure is equivalent to diagonalizing
both sides of \pdj\ and then using the eigenvectors to work
backwards and determine the basis transformation.

The state-operator map 
has been worked out to order $g_2$ in \refs{\basistwo,\psvvv,\gomis}.
In order to write the transformation in a succinct form,
let us denote by $|1,n\rangle$ the normalized state corresponding
to the single-trace
BMN operator $\sum_l e^{2 \pi i n l/J} \Tr(\phi Z^l \psi Z^{J-l})$,
and by $|2,y\rangle$ and $|2,n,y\rangle$ respectively the
normalized states corresponding to the double-trace BMN operators
$\Tr(\phi Z^{J_1}) \Tr(\psi Z^{J-J_1})$ and $\sum_l
e^{2 \pi i n l/J_1} \Tr(\phi Z^l \psi Z^{J_1 - l}) \Tr(Z^{J-J_1})$,
where $y = J_1/J$.
Then we introduce the splitting-joining
operator $\Sigma$ (see \refs{\herman,\vv} for details),
whose action on single-string states is
\eqn\aaa{
\Sigma |1,n\rangle=\sum_{m,y} \sqrt{1-y \over J y} 
{\sin^2(\pi n y) \over \pi^2 (n-m/y)^2} |2,m,y\rangle-
\sum_y
{\sin^2(\pi n y) \over \sqrt{J} \pi^2 n^2} |2,y\rangle.
}
The transformation between the gauge theory basis
$|\psi \rangle$ and the string theory basis $|\widetilde{\psi}\rangle$
may then be written as
\eqn\basisx{
| \widetilde{\psi} \rangle = |\psi\rangle - {g_2 \over 2} \Sigma
|\psi\rangle + {\cal{O}}(g_2^2).
}
As expected, a single-string state corresponds to a single-trace
operator with an order $g_2$ admixture of double-trace operators.
In \psvvv, it was conjectured that the basis transformation
takes the form
\eqn\psvvvbasis{
|\widetilde{\psi}\rangle = e^{-g_2 \Sigma/2} |\psi\rangle,
}
to all orders in $g_2$.

The basis transformation \basisx\ was chosen in part
to ensure that the matrix
elements of \pdj\ agree at order $g_2$, so we need to go to next
order to get a nontrivial check of the BMN correspondence.
The order $g_2^2$ matrix elements of $\Delta - J$ 
in the conjectured string basis $|\widetilde{\psi}\rangle$
given by equation \psvvvbasis\ 
are \refs{\psvvv}
\eqn\mtwo{\eqalign{
\langle \widetilde{1}, m, \pm |(\Delta - J)_{(2)}| \widetilde{1}, n, \pm
\rangle =
{\lambda'g_2^2 \over 16 \pi^2} (B_{mn} - B_{m,-n}).
}}
Here the subscript $(2)$ denotes the order $g_2^2$ part, the matrix
$B_{mn}$ is defined in appendix B (it corresponds to 
non-nearest neighbor interaction diagrams in gauge theory), 
and we have introduced
\eqn\pmdef{
|\widetilde{1},
n, \pm \rangle = {1 \over \sqrt{2}} ( |\widetilde{1},
n \rangle \pm |\widetilde{1}, -n\rangle).
}
The matrix element \mtwo\ vanishes between $|+\rangle$ and $|-\rangle$
states because $|+\rangle$
is in the ${\bf 9}$ representation of SO(4)
while $|-\rangle$ is in the ${\bf 6}$.

Still \mtwo\ does not obviously give a nontrivial check of the BMN
correspondence, because the matrix element is sensitive to the order
$g_2^2$ part of the basis transformation \basisx, and in writing
\mtwo\ we used the conjecture \psvvvbasis\
to order $g_2^2$. 
However,
it was shown in \basistwo\ that the diagonal matrix
elements ($m=n$) are insensitive to the order $g_2^2$ freedom
in adjusting the basis transformation.
This is because an order $g_2^2$ transformation 
in \basisx\ shifts
the matrix elements of \mtwo\ by an amount proportional
to the difference between the order $g_2^0$ energies of the
in- and out-states, which vanishes if $m = \pm n$.
However, the off-diagonal matrix elements in \mtwo\ can be
set to any desired value by choosing the order $g_2^2$ term
in \basisx\ appropriately.
The possibility of performing such transformations
is essentially due to the fact that the gauge theory analog of splitting and 
joining of strings is not corrected at order $g_2^2$.

In summary:  if the string field theory calculation
fails to reproduce the off-diagonal terms in \mtwo, this is
merely an indication that the basis transformation \psvvvbasis\ needs
to be adjusted at order $g_2^2$.
On the other hand, if the string
field theory calculation fails to reproduce the diagonal terms in \mtwo,
then something must be seriously wrong.

\newsec{Light-cone Superstring Field Theory}

In light-cone string field theory
we study the dynamical symmetry generators
$P^-$ and $Q$ expanded in powers of the string coupling $g_s$,
\eqn\expansion{
P^- = P^-_{(0)} + g_s P^-_{(1)} + g_s^2 P^-_{(2)} + \cdots,
\qquad
Q = Q_{(0)} + g_s Q_{(1)} + \cdots.
}
The leading interactions $P^-_{(1)}$ and $Q_{(1)}$, which are cubic
in string fields and mediate simple string joining and splitting, were
determined
in \svone\ for the plane wave background.

In this paper we study $P^-_{(2)}$ with the aim of
reproducing the matrix elements \mtwo\ calculated in the gauge theory.
All of the terms in \expansion\ are determined in principle
by requiring closure of the plane wave supersymmetry algebra
to all orders in $g_s$.
At second order, we have the relation
\eqn\blahtwo{
2P_{(2)}^-  = \{ Q_{(1)}, \overline{Q}{}_{(1)} \}
+  \{ Q_{(0)}, \overline{Q}{}_{(2)} \}
+ \{ Q_{(2)}, \overline{Q}{}_{(0)} \}.
}
Although $Q_{(2)}$ is not known, it does not contribute
to the order $g_s^2$  matrix elements of \blahtwo\ between single-string
states since $Q_{(2)}$ is quartic in string fields at tree level.
Therefore we are only interested in the first term in \blahtwo.
Note that the matrix element between single-string states
has not been computed to order $g_s^2$ even in flat space, although
for a discussion of two-string state matrix elements at this order
see \refs{\gsthree, \gkone \gktwo \gkthree-\grseib}

In  supersymmetric light-cone string field theory
the eigenvalues \evalue\ receive the two
contributions
\eqn\divergence{
{1\over 2} \int_{0}^\infty dT \ \langle\widetilde{1},n| P_{(1)}^-
e^{T(P_{(0)}^- - E_n)} P_{(1)}^- | \widetilde{1},n \rangle
 + {1\over 2} \langle\widetilde{1},n|
\{ Q_{(1)}, \overline{Q}_{(1)} \}
| \widetilde{1},n \rangle,
}
corresponding to the two diagrams in Figure 1. The numerical factor in 
the first term is due to the reflection symmetry of the first diagram in 
that figure, while the numerical factor in the second term arises from 
equation \blahtwo.

In this paper we will calculate the two terms in \divergence\ by
first taking the large $\mu p^+ \alpha'$ limit of $P^-_{(1)}$ and $Q_{(1)}$, and
then including in the sum over intermediate states only a particular
class of states --- namely, those string states which have
only two bosonic excitations (for $P^-_{(1)}$) or one bosonic
and one fermionic excitation (for $Q_{(1)}$).
The order of limits problem is one of the poorly
understood aspects of the BMN correspondence \refs{\ksv}.
This order of limits coincides with the order of limits which
has been used in the gauge theory calculation of \evalue.
Namely, $\Delta-J$ has been diagonalized order by order 
in $\lambda$
within the subspace of two-impurity BMN operators, not
order by order in $\lambda'$ within the space of all
operators.  Therefore, in order to compare with \evalue,
our string theory calculation will include only intermediate
states with two impurities.

\fig{The two light-cone diagrams which contribute at order $g_s^2$
to the eigenvalue \evalue.  The first is the iterated
cubic interaction $P_{(1)}^-$, while the second
is the quadratic contact term induced by $\{Q_{(1)}, Q_{(1)} \}$.}
{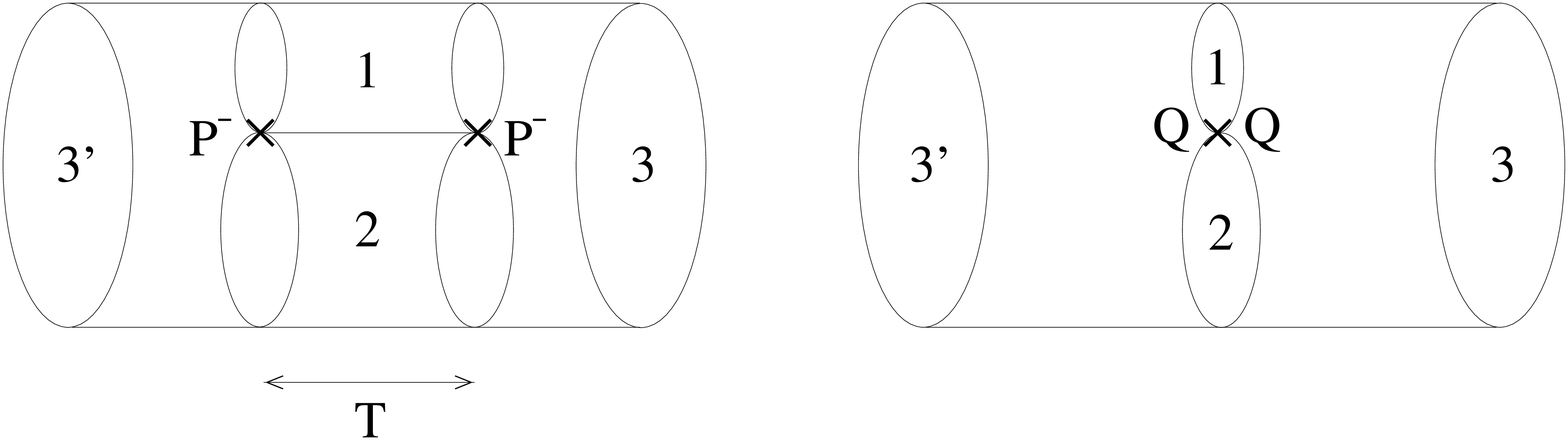}{5.0in}

\subsec{Dynamical generators}

The matrix elements of the dynamical generators in the plane
wave background were presented in \refs{\svone}
to first order in $g_s$ up to an overall function 
$v(\mu, \alpha_3, y),$
where $\alpha_{(r)} = 2 p^+_{(r)}$ and $y =-{\alpha_{(1)}\over
\alpha_{(3)}}$.
They can be expressed as states in the three-string Fock space as

\eqn\qdef{\eqalign{
|P^-\rangle&=\left[(K^I_+ + K^I_-)(K^J_+ - K^J_-)
- \ha \mu \alpha \delta^{IJ}\right] v_{IJ} |V \rangle,\cr
|Q_{\dot{a}} \rangle &= (K_+^I s_{I \dot{a}} + i K_-^I t_{I \dot{a}}) |V
\rangle,\cr
|\overline{Q}_{\dot{a}} \rangle &= (K_+^I t_{I \dot{a}}
- i K_-^I s_{I \dot{a}} ) |V\rangle.
}}
We refer the reader to the papers \refs{\svone,\svtwo,\ari}
for details, since we will present only the parts of \qdef\ which
are relevant to our calculation.

The vertex $|V\rangle$ is given by
\eqn\aaa{
|V\rangle= \exp \left[ \ha \sum_{r,s=1}^3
\sum_{m,n=-\infty}^\infty a^{I \dagger}_{m(r)}
\overline{N}^{(rs)}_{mn} a^{J \dagger}_{n(s)} \delta_{IJ} 
\right] E_b E_b^0|0\rangle|0\rangle|0\rangle,
}
where $E_b$ is exponential in fermionic creation operators
\eqn\aaa{
E_b=\exp\left[\sum_{r,s=1}^3 \sum_{m,n=1}^\infty
b^\dagger_{-m(r)} Q^{(rs)}_{mn} b^\dagger_{n(s)}-\sqrt{2}
\Lambda \sum_{r=1}^3 \sum_{m=1}^\infty Q^{(r)}_m b^\dagger_{-m(r)}
\right]
}
and $E_b^0$ is linear
in the fermionic zero modes,
\eqn\aaa{
\qquad E_b^0=  \prod_{a=1}^8 (\lambda^a_{(1)} + 
\lambda_{(2)}^a + \lambda_{(3)}^a).
}
The tensors $s,$ $t$ and $v$ are defined by
\eqn\aaa{
\eqalign{
v^{IJ}&= \delta^{IJ} - {i \over \alpha} \gamma^{IJ}_{ab}
Y^a Y^b + {1 \over 6 \alpha^2}
\gamma^{IK}_{ab} \gamma^{JK}_{cd} Y^a Y^b Y^c
Y^d\cr
&~- {4 i\over 6! \alpha^3} \gamma^{IJ}_{ab}
\epsilon_{abcdefgh} Y^c Y^d Y^e Y^f
Y^g Y^h
+ {16 \over 8! \alpha^4} \delta^{IJ}
\epsilon_{abcdefgh}
Y^a Y^b Y^c Y^d Y^e Y^f
Y^g Y^h,\cr
s_{I\dot{a}} &= 2 \gamma^I_{\dot{a} a} Y^a - {8 \over 6!
\alpha^2} \gamma^{IJ}_{ab} \gamma^J_{c \dot{a}} \epsilon^{abcdefgh}
Y^d Y^e Y^f Y^g Y^h,\cr
t_{I \dot{a}} &= -{2 \over 3 \alpha}
\gamma^{IJ}_{ab} \gamma^J_{c \dot{a}} Y^a Y^b Y^c
- {16 \over 7! \alpha^3} \gamma^I_{a \dot{a}} \epsilon^{abcdefgh}
Y^b Y^c Y^d Y^e Y^f Y^g Y^h,
}}
where $\alpha = \alpha_{(1)} \alpha_{(2)} \alpha_{(3)}$.
Finally, the quantities $K_\pm$ and $Y$ are linear in creation
operators,
\eqn\aaa{\eqalign{
K^I_+&=\sum_{r=1}^3  \sum_{m=0}^{\infty} F^+_{m(r)} a^{I \dagger}_{m(r)},\qquad
K^I_-=\sum_{r=1}^3  \sum_{m=1}^{\infty} F^-_{m(r)} a^{I \dagger}_{-m(r)},\cr
Y&=Y_0
(\alpha_{(1)}
\lambda_{(2)}-
\alpha_{(2)} \lambda_{(1)}) +
\sum_{r=1}^3 \sum_{m=1}^{\infty} Y_{m(r)} b^{\dagger}_{m(r)}.
}}

\subsec{Dynamical generators at large $\mu p^+ \alpha'$}

The matrix elements $\overline{N}^{(rs)}_{mn}$, $F^\pm_{m(r)},$
$Y_{m(r)},$ $Q^{(rs)}_{mn}$ and $Q^{(r)}_m$ 
are known for all $\mu p^+ \alpha'$.
In this subsection we write the leading behavior of 
these quantities for large $\mu p^+_{(3)} \alpha'$.
In this limit the nonzero Neumann matrices are
\foot{We set $\alpha'$=2 here.}
\eqn\none{
\overline{N}^{(13)}_{mn}=(-1)^{m+n}{2 \over \pi}
{y^{3\over 2} n \sin{\pi n y} \over {m^2-n^2 y^2}},
\qquad
\overline{N}^{(23)}_{mn}=(-1)^{n+1} {2 \over \pi}
{(1-y)^{3\over 2} n \sin{\pi n y} \over {m^2-n^2 (1-y)^2}},
}
\eqn\ntwo{
\overline{N}^{(13)}_{0n}=(-1)^{n+1} {1 \over \pi} \sqrt{2 \over y}
{\sin{\pi n y} \over n},
\qquad
\overline{N}^{(23)}_{0n}=(-1)^{n} {1 \over \pi} \sqrt{2 \over 1-y}
{\sin{\pi n y} \over n},
}
and
\eqn\nthree{
\overline{N}^{(r3)}_{-m-n}={m \over n} {\alpha_{(3)} \over \alpha_{(r)}} 
\overline{N}^{(r3)}_{mn},
}
where $m,n > 0$ and $y=-{\alpha_{(1)} \over \alpha_{(3)}.}$
The fermionic Neumann matrices can be obtained similarly.

The leading matrix elements of the prefactor which will appear
in the calculations of the next section are

\eqn\aaa{\eqalign{
F_{n(3)}^-&= i (-1)^{n+1} \sqrt{2 \alpha_{(1)} \alpha_{(2)} \over \pi}
\sin\pi n y,\cr
Y_{m(1)}&= {1 + \Pi \over 2} 
\sqrt{\alpha_{(2)} |\alpha_{(3)}| \over 4 \mu \pi} 
(-1)^{m+1} ,\cr
Y_{m(2)}&= {1 + \Pi \over 2} 
\sqrt{\alpha_{(1)} |\alpha_{(3)}| \over 4 \mu \pi},\cr
Y_0 &= \sqrt{|\alpha_{(3)}| \over 4 \mu \pi \alpha_{(1)} \alpha_{(2)}}
\left[4 \pi \mu  {\alpha_{(1)} \alpha_{(2)}
\over|\alpha_{(3)}|} {1 - \Pi \over 2} + {1 + \Pi \over 2}\right].
}}
For other bosonic components see \svtwo, \hssv, and for fermionic \ari.

\subsec{Matrix element of $(P_{(1)}^-)^2$}

The first contribution to the eigenvalue \evalue\ comes from
the iterated $P_{(1)}^-$ interaction (the first term
in \divergence).
This calculation was performed in \psvvv\ using the expressions
for $P_{(1)}^-$ at large $\mu p^+ \alpha'$ 
given in the previous subsection,
with the result
\eqn\hsquared{
{2 \over \mu}
\left[ {1\over 2} \langle \widetilde{1}, n, + | P_{(1)}^- {1\over \Delta}
P_{(1)}^- | \widetilde{1}, n, + \rangle \right]=
{1 \over \mu}
\sum_i { \langle \widetilde{1}, n, + | P_{(1)}^- |i\rangle \langle i|
P_{(1)}^- | \widetilde{1}, n, + \rangle
\over E_n - E_i} = - {g_2^2 \lambda' \over 8 \pi^2} B_{n,-n},
}
where, as discussed above, the sum runs over intermediate
2-string states with only two bosonic excitations (which can either be on the same
string or different strings).
Note that the matrix element on the right-hand side
of \hsquared\ is expressed in terms of unit-normalized
states (which are natural from the gauge theory perspective)
rather than delta-function normalized
states (see \seven). In writing the final result we chose to normalize 
$P^-_{(1)}$ to absorb a number of irrelevant factors
by taking $v(\mu, \alpha_{(3)}, y)=
g_2/(\langle v^{II}\rangle 2 \sqrt{2} \alpha_{(3)}^3).$
Finally, note that the analogous calculation \hsquared\ for
the state $|\widetilde{1}, n, -\rangle$ gives zero, since the
matrix element of $P^-_{(1)}$ between $|\widetilde{1}, n, -\rangle$
and any two-impurity two-string state vanishes.

\subsec{Matrix element of $\{\overline{Q}_{(1)},
Q_{(1)}\}$}

In this subsection we will calculate the matrix element
$\{\overline{Q}_{\dot a},Q_{\dot b}\}$ (suppressing
the subscript ${(1)}$) for an analogous
class of intermediate states: those 2-string states
which have only one bosonic excitation and one fermionic
excitation.  
There are two possibilities.  If the worldsheet mode number $m$
is nonzero, then the impurities have to sit on the same string,
so in the conventions of \svone\ we have
\eqn\aaa{
|1 \rangle = {1 \over 2} (a_{m}^{K \dagger} + i e(m)
a_{-m}^{K \dagger})( b_{m}^{a \dagger} - i e(m) b_{-m}^{a \dagger})
|{\rm v}\rangle, \qquad |2 \rangle = |{\rm v}\rangle,
}
where $e(m) = {\rm sign}(m)$, $|{\rm v}\rangle$
is the ground-state of the world-sheet Hamiltonian,
and $K$ and $a$ are spinor indices that
will be summed over all possible values for the intermediate states.
When $m=0$, the two oscillators $a_{0}^{K \dagger}$
and $b_{0}^{a \dagger}$ can sit on either the same string or different
strings, and all possibilities are included in the sum over intermediate
states.

The single-string states introduced in \pmdef\ may be written
with the string oscillator conventions of \svone\ as
\eqn\statesother{\eqalign{
|{\tilde 1},n,+ \rangle={1 \over \sqrt{2}}
(a^{I \dagger}_{n} a^{J \dagger}_n+a^{I \dagger}_{-n}
a^{J \dagger}_{-n})|{\rm v} \rangle,\cr
|{\tilde 1},n,- \rangle={1 \over \sqrt{2}}
(a^{I \dagger}_{n} a^{J \dagger}_{-n}-a^{I \dagger}_{-n}
a^{J \dagger}_{n})|{\rm v} \rangle.
}}

It can be checked that only the terms with $K_-$ in 
$Q$ and $\overline{Q}$ 
are leading at large $\mu p^+ \alpha'.$
Moreover only the term $Y^5$ contributes in $s_I$
and term $Y^3$ contributes in $t_I.$

With these clarifications, it is not hard to construct the  matrix elements 
of the anticommutator $\{ \overline{Q}_{\dot a},{Q}_{\dot b}\}$,
\eqn\pnpn{
\langle \tilde{1}, n,+|\{ \overline{Q}_{\dot a},{Q}_{\dot b}\}
|\tilde{1},p,+ \rangle=
{g_2^2 \delta_{{\dot a} {\dot b}} \over 8 \alpha_{(3)}^6}
\int_{0}^1 {dy \over y (1-y)}
(F^-_{n(3)})^2 \sum_{m=1}^{\infty}
\sum_{s=1,2} \overline{N}^{(s3)}_{-m -n} \overline{N}^{(s3)}_{-m -p} (Y_{m(s)})^2
}
\eqn\mnmn{
\eqalign{
&\langle \tilde{1},n,-|\{\overline{Q}_{\dot a},Q_{\dot b}\}
|\tilde{1},p,- \rangle=
{g_2^2 \delta_{{\dot a} {\dot b}}\over 8 \alpha_{(3)}^6 }
  \int_{0}^1 {dy \over y (1-y)} 
 (F^-_{n(3)})^2
\cr
&\times
\left[
\sum_{s=1,2}
\sum_{m=1}^\infty \overline{N}^{(s3)}_{mn}  \overline{N}^{(s3)}_{mp}(Y_{m(s)})^2+2
\sum_{s,s'=1,2} { \alpha_{(3)} y^2 (1-y)^2 \over \alpha_{(s')}}
\overline{N}^{(s3)}_{0n}\overline{N}^{(s3)}_{0p}  
(Y_{0(s')})^2\right].
}}
It is not hard to see the origin of the various factors appearing
in these equations.  In \pnpn, the bosonic excitation on the
intermediate string must have negative mode number (in our basis,
negative and non-negative modes do not couple).
In \mnmn, the second term involves a double sum over
both strings because when the intermediate excitations are zero-modes
they do not have to sit on the same string.
Finally, the measure arises because in computing
the matrix element above we inserted a complete set of
(physical) 2-string states (with two excitations) between 
$Q$ and ${\overline{Q}}$. We used the string theoretic normalization of states, $\langle p^+_i|
p^+_j\rangle=p^+_i\delta(p_i^+-p_j^+)$. Thus, the term
$|i\rangle|j\rangle
\langle i|\langle j|$ in the identity operator for the 2-string states appears multiplied 
by $1/(p^+_ip^+_j)$, together with an integral for each of the two momenta. 
Because of the $p^+$ conservation constraint for each matrix element one of the 
two $p^+$ integrations can be trivially performed and one is left with only one 
integral, with the measure displayed above. Finally, the overall coefficient 
arises due to certain relations between $v^{IJ}$ and the derivative 
of $s^{I}_{\dot a}$ with respect to $Y$ \svone.

Using the sums from the appendix as well as the choice of normalization of $P^-_{(1)}$, 
the expressions
for $Y_m$ and $F_m$ from the previous section, we find for the diagonal pieces
\eqn\qqres{\eqalign{
{2 \over \mu}
\langle \tilde{1},n,+|\{\overline{Q}_{\dot a},
Q_{\dot b}\}|\tilde{1},n,+ \rangle=
 {g_2^2 \lambda' \over 8 \pi^2} \delta_{{\dot a}{\dot b}} ({1\over 3}+{5 \over 8 \pi^2 n^2})=
 {g_2^2 \lambda' \over 8 \pi^2} \delta_{{\dot a}{\dot b}} (B_{nn}+B_{n,-n}),
\cr
{2 \over \mu} \langle \tilde{1},n,-|\{\overline{Q}_{\dot a},Q_{\dot b}\}
|\tilde{1},n,- \rangle=
{g_2^2 \lambda' \over 8 \pi^2} \delta_{{\dot a}{\dot b}} ({1\over 3}+{35 \over 8 \pi^2 n^2})=
{g_2^2 \lambda' \over 8 \pi^2} \delta_{{\dot a}{\dot b}} (B_{nn}-B_{n,-n}).
}}
Extracting the matrix elements of $P^-$ and adding them to \hsquared, we find that the total matrix 
element matches
the gauge theory result of \refs{\basistwo,\basisone,\basisthree}.
Note that we have not attempted to match the overall coefficient,
since the precise normalization of \qdef\ in string field theory
has not yet been fixed.   The relative normalization
between \hsquared\ and \qqres\ is the same due to supersymmetry \psvvv.
The fact that
the $n$ dependence of \qqres\ works out is highly nontrivial.
It is also clear that the normalization of $Q$ cannot be independently 
adjusted, since it is related to the normalization of $H$ by the order $g_s$ 
supersymmetry algebra.

For the off-diagonal transition 
the result of the summation and integrations over $y$ is
\eqn\offdiag{
\eqalign{
{2 \over \mu} \langle \tilde{1}, n,+|\{{\overline{Q}}_{\dot a},\,{Q}_{\dot b}\} 
|\tilde{1},p,+\rangle&={g_2^2 \lambda' \over 8 \pi^2} \delta_{{\dot a}{\dot b}}
\left[ {n^2+p^2\over \pi^2 (n^2-p^2)^2}\right]=
{g_2^2 \lambda' \over 8 \pi^2} \delta_{{\dot a}{\dot b}} (B_{np}+B_{n,-p}),\cr
{2 \over \mu}
\langle \tilde{1},n,-|\{{\overline{Q}}_{\dot a},\,{Q}_{\dot b}\} |\tilde{1},p,-\rangle&=
{g_2^2 \lambda' \over 8 \pi^2} \delta_{{\dot a}{\dot b}}
\left[{3n^4-4n^2p^2+3p^4\over \pi^2 n p (n^2-p^2)^2}\right]=
{g_2^2 \lambda' \over 8 \pi^2} \delta_{{\dot a}{\dot b}} (B_{np}-B_{n,-p}).
}}
This expression agrees with the off-diagonal elements
in \mtwo, giving the further support to the conjecture 
of \psvvv.

\newsec{Discussion and Conclusion}

In this note we have calculated order $g_2^2$ matrix elements
of the string field theory Hamiltonian between single-string states.
The diagonal matrix elements
are precisely those needed to recover the eigenvalues \evalue\ of
$\Delta-J$ which have been computed on the gauge theory side
within the subspace of two-impurity BMN operators.
Moreover, we find agreement between the off-diagonal
matrix elements \offdiag\ and those predicted by the conjectured
state-operator map \psvvvbasis, 
thereby confirming the proposal of \psvvv\ to order $g_2^2$.

It is important to stress that the eigenvalues \evalue\ 
result from a truncated calculation on the gauge theory side.
Specifically, the operator $\Delta - J$ has been
diagonalized perturbatively, order by order in $\lambda$,
within the subspace of two-impurity BMN operators.
It has been stressed in \refs{\ksv, \psvvv} that there is
no reason why the large $J$ limit of the $\lambda$
perturbation series has to agree order by order with
the $\lambda'$ series.  (Recall that the BMN limit
requires taking the $\lambda \to \infty$ limit of
quantities which can be calculated in the gauge
theory only at small $\lambda$.)
Therefore, in order to compare with the result \evalue, we
have done a similar truncation of the string field theory
calculation, by including only two-impurity intermediate
states.

Matrix elements in which the number of impurities are not conserved
have not yet been analyzed in the gauge theory, but on the string
theory side it was pointed out in \svtwo\ that matrix elements
in which two impurities are created or destroyed are actually
larger, by a factor of $\mu p^+ \alpha'$, than impurity-conserving matrix elements.
Since $\mu p^+ \alpha'= 1/\sqrt{\lambda'}$, it seems that these matrix elements 
cannot be seen in perturbation theory around $\lambda' = 0$. This would 
indeed be the case if $\lambda'$ were the coupling constant order by order 
in perturbation theory. While this seems to be true for operators with
$\Delta - J=1$ impurities, it may happen that at some higher order these 
operators mix with ones with $\Delta - J\ge 2$ impurities. Then, perturbation 
theory will become an expansion in $\lambda$ rather than $\lambda'$. Since the 
't~Hooft coupling is taken to be large, reliable results require resummation 
of the perturbation theory. Then, the appearance of $1/\sqrt{\lambda'}$ becomes
a strong coupling effect, similar to the appearance of $\sqrt{\lambda}$ in the 
Wilson loop calculations in the context of the AdS/CFT correspondence. Unlike that 
case the eventual emergence of $1/\sqrt{\lambda'}$  should be signaled in perturbation 
theory by a divergence in the limit $J\rightarrow \infty$ at some loop order.

Both terms in \divergence\ receive contributions from intermediate states with
more than two creation operators. In the large $\mu p^+ \alpha'$ limit, the only 
non-vanishing matrix elements in which the number of impurities are not conserved
require changing the number of impurities by two. Each matrix element of this type 
is larger by a factor of $\mu p^+ \alpha'$ 
compared to matrix elements in which the number 
of impurities is conserved. However, the energy denominator is also larger by a factor
of $(\mu p^+ \alpha')^2$ and thus these intermediate states contribute to leading order. 
It turns out that the contribution of these states is actually linearly divergent.
This is related to the fact that the computation is done in the large 
$\mu p^+ \alpha'$ limit. 
At finite $\mu p^+ \alpha'$ this divergence is regularized. 
Since the initial divergence was linear, 
taking the large $\mu p^+ \alpha'$ 
limit at the end leads to a contribution of order 
$\mu p^+ \alpha'$ to 
the masses of string states. The gauge theory counterpart of this effect is a 
${1\over\sqrt{\lambda'}}$ contribution to the anomalous dimension of some (appropriately 
redefined) operators. 

It remains a very interesting outstanding problem to go beyond
the truncation to two-impurity intermediate
states on either side of the plane-wave gauge/string theory
duality.  On the gauge theory side, this would apparently require
diagonalizing $\Delta - J$ at finite $\lambda$ within the space
of all gauge theory operators, and then taking the $\lambda,J \to
\infty$ limits to decouple the non-BMN operators.

\bigskip

\centerline {\bf Acknowledgements}

\smallskip

We are grateful to
S. Giddings,
D. Gross, D. Freedman, M. Headrick, 
S. Minwalla, L. Motl, M. Ro\v{c}ek, J. Schwarz,
D. Vaman and
H. Verlinde for useful discussions and comments.
R. R. is supported in part by DOE under 
Grant No. 91ER40618 and in part by NSF 
under Grant No. PHY00-98395.
M.S. is supported by DOE under Grant No. DE-FG02-91ER40671, and
A.V. is supported by NSF under Grant No. PHY99-07949.
Any opinions, findings, and conclusions
or recommendations expressed in this material
are those of the authors and do not necessarily reflect the views
of the National Science Foundation.

\appendix{A}{Useful Sums}

For $n,p>0$ and $r \in \{1,2\}$ we obtain
from \none, \ntwo\ and \nthree\ the identities
\eqn\aaa{
\eqalign{
\sum_{q\ge 0}\overline{N}^{(3s)}_{nq}
\overline{N}^{(3s)}_{pq}&=-{(\delta_{s,1} (-1)^{n+p} +\delta_{s,2}) 
\over \pi}
\left[{\sin\left(\pi (n-p){\alpha_{(s)} \over \alpha_{(3)}}\right)  
\over n-p}
+{\sin \left(\pi (n+p) { \alpha_{(s)} \over \alpha_{(3)}}\right)
\over n+p}
\right],\cr
\sum_{q> 0}\overline{N}^{3s}_{-n-q}\overline{N}^{(3s)}_{-p-q}&=
-{(\delta_{s,1} (-1)^{n+p} +\delta_{s,2})\over \pi}\left[
{\sin\left(\pi (n-p){\alpha_{(s)} \over \alpha_{(3)}}\right)
\over n-p} -
{\sin \left(\pi (n+p) { \alpha_{(s)} \over \alpha_{(3)}}\right)
\over n+p}
\right].
}}

\appendix{B}{The Matrix $B$}

The matrix $B$ is given by \seven
\eqn\coefb{
B_{nm} = \cases{
0 & n=0 if $m=0$ \cr
\frac13+\frac{10}{u^2} & if $n=m\neq0 $ \cr
-\frac{15}{2u^2} & if $n=-m\neq0 $ \cr
\frac6{uv}+\frac2{(u-v)^2} & {all other cases,} \cr }}
where $u=2 \pi m$, $v=2 \pi n$.

\listrefs
\end